\documentclass{aa}

\usepackage{natbib,twoopt}
\usepackage{graphicx}
\usepackage{array}
\usepackage{txfonts}
\usepackage{soul}
\usepackage{xcolor}
\usepackage{caption}
\usepackage{longtable}
\usepackage{bigstrut}
\usepackage{multirow}
\usepackage{hyperref}

\newcommand{\tstrut}{\rule{0pt}{2.6ex}}
\newcommand{\bstrut}{\rule[-0.9ex]{0pt}{0pt}}

\begin{document} 

   \title{Molecular gas in the disks and tails of ram-pressure-stripped Coma~galaxies}
   
    \author{A.~Srbljanović\inst{1,2}\corrauth{kabatova@asu.cas.cz}
    \and P.~Jáchym\inst{2}\email{jachym@ig.cas.cz}
    \and M.~Sun\inst{3}\email{ms0071@uah.edu}
    \and J.~Kenney\inst{4}\email{jeff.kenney@yale.edu}
    \and L.~Cortese\inst{5}\email{luca.cortese@uwa.edu.au}
    \and F.~Combes\inst{6}\email{francoise.combes@obspm.fr}
    \and M.~Yagi\inst{7}\email{yagi.masafumi@nao.ac.jp}
    \and J.~Palouš\inst{2}\email{palous@ig.cas.cz}
    \and A.~Anitha~Shaji\inst{2}\email{aashiya.shaji@asu.cas.cz}
    \and E.~Roediger\inst{8}\email{E.Roediger@hull.ac.uk}
    \and B.~Deshev\inst{9}\email{boris.deshev@ut.ee}}

   \institute{
   Faculty of Nuclear Sciences and Physical Engineering, Czech Technical University in Prague, Břehová 7, 11000 Prague, Czechia
   \and
   Astronomical Institute of the Czech Academy of Sciences, Boční II 1401, 14100 Prague, Czechia
   \and
   Department of Physics and Astronomy, University of Alabama in Huntsville, 301 Sparkman Dr NW, Huntsville, AL 35899, USA
   \and
   Department of Astronomy, Yale University, New Haven, CT 06511, USA
   \and
   International Centre for Radio Astronomy Research (ICRAR), University of Western Australia, Crawley, WA 6009, Australia
   \and
   Observatoire de Paris, LERMA, Collège de France, CNRS, PSL University, Sorbonne University, 75014 Paris, France
   \and
   National Astronomical Observatory of Japan, Osawa 2-21-1, Mitaka, Tokyo 181-8588 Japan
   \and
   E.A. Milne Centre, University of Hull, Cottingham Road, HU6 7RX, U.K.
   \and
   Tartu Observatory, Faculty of Science and Technology, University of Tartu, Estonia
}

   \date{Received, Accepted}
   
\abstract{
We present new IRAM 30\,m observations of the CO(1--0) and CO(2--1) emission in five galaxies of the Coma cluster undergoing ram pressure stripping. Molecular gas is detected in four galaxies and in three cases, we clearly detect CO emission in the stripped tails, extending up to $\sim$50~kpc from the stellar disks. The molecular gas masses in the tails start from a few $\times 10^{8}\,M_{\odot}$ and account for up to 10--20\% of the molecular gas remaining in the parent disks.\par
When compared to field scaling relations, several galaxies exhibit elevated molecular-to-stellar gas ratios, despite being \ion{H}{i} deficient. Notably, this enhancement persists toward the lower-mass end of our sample, even in the massive Coma cluster environment. This is consistent with ram pressure-induced compression of atomic gas and an enhanced \ion{H}{i}-to-H$_2$ conversion in the disks. At the same time, the total gas fractions of most galaxies remain close to those of field galaxies of similar stellar mass once the molecular gas in the tails is included, indicating that gas removal and phase transformation are simultaneously at work.\par 
We further investigate the relation between molecular gas and ionized gas by comparing CO and H$\alpha$ emission in the stripped regions. While CO is preferentially detected in H$\alpha$-bright regions, the increased scatter compared to previous studies suggests the presence of both star-forming and diffuse CO gas components in the tails. Our results highlight the dual role of ram pressure stripping in dense environments: it efficiently removes gas from galaxies while also promoting molecular gas formation both within the disks and in the stripped material, shaping the short-term star formation phase in cluster galaxies.
}


\maketitle
\nolinenumbers

\section{Introduction}
The hot and dense environment of galaxy clusters represents a site of dramatic changes in the morphology of member galaxies as they evolve from blue gas-rich to red gas-poor types due to multiple environmental processes. Apart from internal processes influencing the gas reservoir of galaxies, such as winds originating both from star formation and active galactic nucleus (AGN) accretion \citep{Veilleux2005, Croton2006}, there are important environment-dependent processes acting from the outside as well. Gravitational (or tidal) interactions can occur between two galaxies \citep{Toomre1972} or between the galaxy and the cluster potential \citep{Byrd1990}. Moreover, the combined effect of a large number of high-velocity galaxy-galaxy encounters is referred to as harassment \citep{Moore1996}. These mechanisms affect all components of the galaxies including the stellar disk which is the most prominent difference between them and the hydrodynamical interactions that leave the stellar disk unperturbed. These involve thermal evaporation \citep{Cowie1977}, viscous stripping \citep{Nulsen1982}, and ram pressure stripping (RPS) first described by \citet{Gunn1972}. Irrespective of the main driver of the gas removal, the star formation is quenched only if the galaxy also loses the ability to accrete fresh gas from the circumgalactic medium (CGM) known as starvation or strangulation \citep{Larson1980}.\par

One of the key processes of the environmental galaxy quenching is RPS experienced by galaxies as they move at high velocities through the hot intracluster medium (ICM), whose temperatures reach $\sim$10$^7$ -- $\sim$10$^8$ K. During this process, the gas reservoir of galaxies is gradually depleted from outside in as suggested by many observations of partially or fully gas-stripped galaxies \citep{Cayatte1990,BravoAlfaro2001,Chung2009,Loni2021}. Without any gas left in the disk, the star formation  processes of the galaxy are quenched. However, a very different story evolves in the tail behind the parent galaxy. After leaving the disk, the displaced gas presumably mixes with the ICM forming a complex environment whose temperature and density depend locally on prevalence of heating and cooling processes. However, details of the ISM evolution remain unknown and are the subject of the intense research; among many: the observational surveys ALMA-JELLY (Jáchym et al. in prep.; \citealt{Souchereau2025}), GASP \citep{Poggianti2017}, VIVA \citep{Chung2009}, VESTIGE \citep{Boselli2018}, and the ALMA Fornax Cluster Survey \citep{Zabel2019}; along with simulations and theoretical works \citep{Yun2018, Tonnesen2012, Roediger2009}.\par

While the \ion{H}{i} gas is stripped effectively, the depletion of molecular gas in giant molecular clouds (GMCs) via the ram pressure stripping is rare; presumably, this is because they are typically denser and located closer to the galaxy center. Observationally, the amount of H$_2$ in the cluster galaxies has been found to be similar to the one of field galaxies or slightly lower \citep{Cortese2021}. This deficiency could be accounted for by the depletion of the diffuse \ion{H}{i}, which fuels GMCs and does not have to be a consequence of the direct stripping. However, under extreme conditions, the direct stripping may be possible even for dense GMCs \citep{Farber2021}. On the other hand, recent work of \citet{Moretti2020} shows excess of the molecular gas in cluster galaxies compared to the sample of field galaxies, possibly explained by an effective conversion from \ion{H}{i} to H$_2$ enhanced by the ongoing ram pressure-stripping.\par
As for the molecular gas in tails, simulations suggest that new GMCs can be formed in situ from the stripped gas mixed with the ICM, which cools and condensates \citep{Farber2021, Tonnesen2012}. This is likely to explain the presence of young stars in tails completely detached from the original galaxy \citep{Kenney2014, George2018}.\\ \indent
While a number of studies have described one-sided tails of cluster galaxies in multiple wavelengths (a full list given in \citealt{Boselli2021}), observational evidence of this kind of abundance of the cold molecular gas was absent for a long time, even though it was expected due to the confirmed tail star formation. A cold molecular gas tail component was discovered for the first time in the galaxy ESO~137-001 residing in the Norma cluster \citep{Jachym2014}, with an unexpectedly large abundance of CO $\sim$10$^9$\(M_\odot\); result comparable to that of the hot ionized phase of the tail gas. Since then, few more observations have broadened the list: galaxy NGC 4388 in Virgo cluster \citep{Verdugo2015}, galaxy D100 with a prominent 50 kpc-long tail in Coma cluster \citep{Jachym2017}, four CO-rich tails in galaxies JO201, JO204, JO206, and JW100 \citep{Moretti2018}, and (most recently) NGC 4858 in the Coma cluster \citep{Souchereau2025}.\par

This paper follows on previous searches for molecular gas in the tails of Coma cluster members with a sample of five galaxies. We describe new IRAM 30m observations and detection of the molecular gas component both in disks and in tails and estimate its mass. For non-detection, upper limits are estimated. The paper is organized as follows. In Sect. \ref{section_observation} our sample and IRAM observations are introduced. The results are described in Sect. \ref{section_results}. This includes general features of the detections and a description of the subsequent analysis. In Sect. \ref{section_expectation}, the measured molecular gas fraction is compared to expected value from main sequence field galaxies as well as observations of different cluster galaxies on higher redshifts. Moreover, a comparison for total gas fraction and a molecular gas mass to atomic gas mass is shown. In Sect. \ref{section_history}, the galaxies' evolution within the cluster is discussed based on their H$\alpha$ and H$_2$ properties. The relationship of different gas phases is further studied in Sect. \ref{section_halpha}, where a previously described correlation of H$\alpha$ with CO is populated with the data from our sample. Conclusions are provided in Sect. \ref{section_conclusion}.\par 
We use the Coma cluster luminosity distance of $D_L$=102.2 Mpc, angular diameter distance of $D_A$=97.5 Mpc (1 arcsec $\approx$ 0.473 kpc), and redshift z=0.0231. For velocities, radio definition $v$~=~$cz/(z+1)$ is used. 

\section{Data}
\label{section_observation}
 \subsection{Sample of Coma galaxies}
\begin{figure}[!ht]
   \centering
   \includegraphics[width=\hsize]{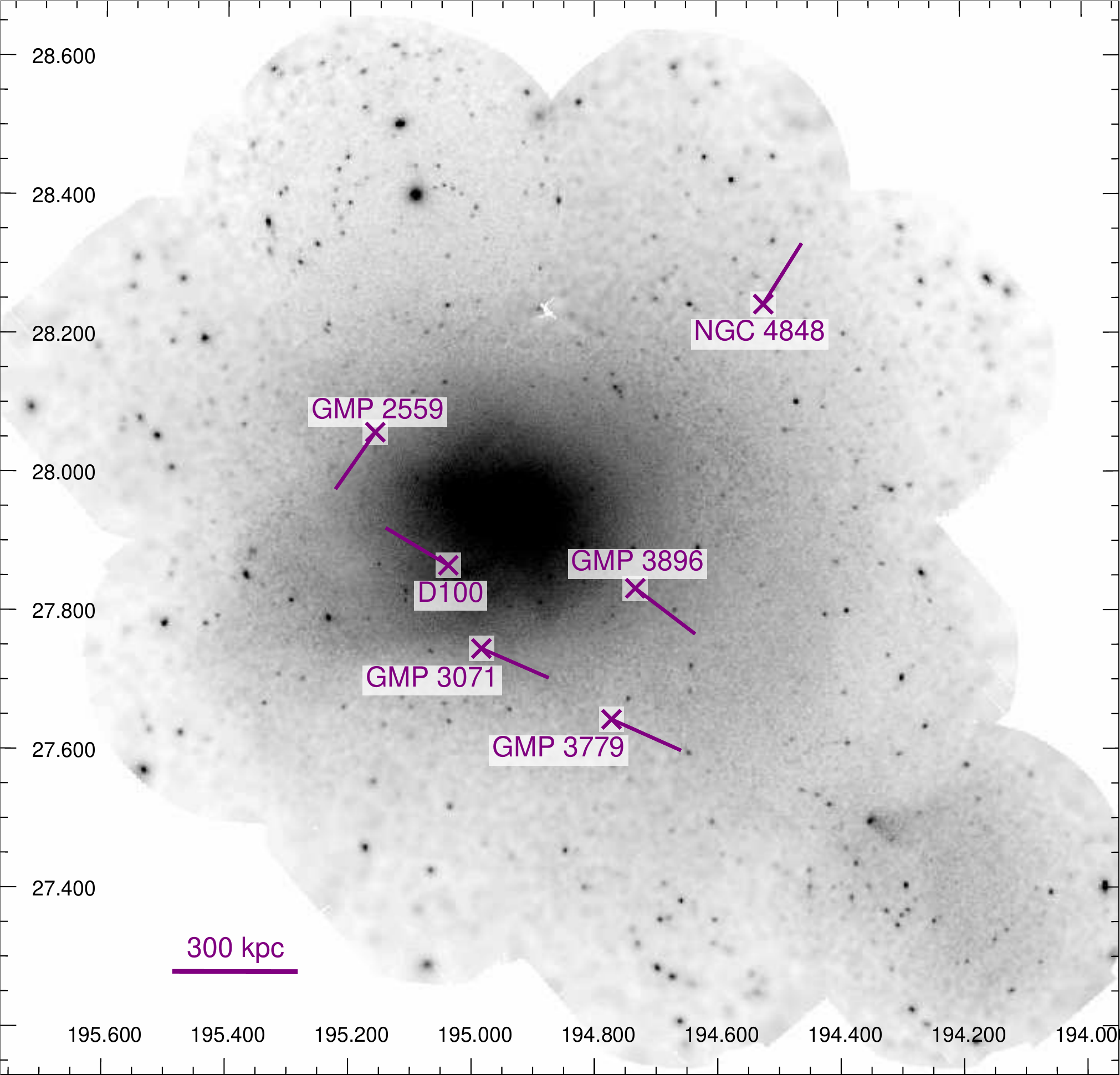}
      \caption{\textit{XMM-Newton} 0.4 - 1.3 keV image of the Coma cluster \citep{Snowden2008} with the locations of sample galaxies marked with purple crosses and indicated tail directions.}
         \label{fig_coma_our_galaxies}
   \end{figure}

The Coma cluster is the nearest ($D_L$ $\sim$ 100 Mpc) rich ($>$1000 members) massive ($M_{dyn}$ $\sim$10$^{15}$\(M_\odot\)) cluster of galaxies. For these reasons, it is an extraordinary laboratory for studying galaxy evolution. The sample includes five Coma RPS galaxies. All of the galaxies are part of the central region of the cluster well within the virial radius of $\sim$2 Mpc, as can be seen in Fig. \ref{fig_coma_our_galaxies}. The sample was designed to follow galaxies in different phases of their evolution within the Coma cluster, as estimated based on their H$\alpha$ properties according to \citet{Yagi2010}: early-phase stripping with bright H$\alpha$ regions connected to star-forming disk; or advanced stripping phase with H$\alpha$ regions connected to the star-forming core, while the extended disk star formation is absent. All of the galaxies show clear evidence of ongoing stripping, including extended one-sided gaseous H$\alpha$ tails, asymmetrical \ion{H}{i} distributions, and significant \ion{H}{i} deficiencies (ranging from 0.25 to 1.01). In Table \ref{table_objects}, we list the properties of observed galaxies.

In addition to the sample presented here, we also used observations of the Coma galaxy D100 (GMP 2910, MRK60) from \citet{Jachym2017}. This galaxy is an extraordinary example of an RPS galaxy and its tail has the biggest length-to-width ratio of all known RPS galaxies. Moreover, it was revealed that the tail is extremely gas-rich with the recovered molecular gas mass of $\sim$10$^9$\(M_\odot\) indicating that the molecular component is actually dominating the tail mass. In this paper, we used this galaxy for comparison in several places because of its resemblance to our sample. 

\begin{table*}
\caption{Properties of observed Coma cluster galaxies.}             
\label{table_objects}      
\centering          
\begin{tabular}{l l c c c c c c c r}
\hline\hline       
Name & Other name & \vtop{\hbox{\strut RA, Dec}\hbox{\strut (J2000) [deg]}} &  \vtop{\hbox{\strut $i$} \hbox{\strut [deg]}} & $z$ & \vtop{\hbox{\strut $V_{\mathrm{Coma}}$}\hbox{\strut [km s$^{-1}$]}} & \vtop{\hbox{\strut $M_\star$}\hbox{\strut [10$^8$ M$_{\odot}$]\bstrut}}  &  \vtop{\hbox{\strut $d_{\mathrm{Coma}}$}\hbox{\strut [kpc]}} & Def$_{\mathrm{\ion{H}{i}}}$ & \vtop{\hbox{\strut Stripping}\hbox{\strut phase}}\\ 
\hline

NGC~4848 & \vtop{\hbox{\strut GMP4471}\hbox{\strut CGCG 160-055}} & 194.523520, 28.242740 & 71 & 0.02351 & 117 & 509.1  & 785 & 0.25 & early\\

GMP~3896 & \vtop{\hbox{\strut IC3949, D89}\hbox{\strut CGCG 160-212}} & 194.733375, 27.833361 & 81 & 0.02525 & 615 & 345.3  & 415 & 1.01 & early\\ 

GMP~2559 & \vtop{\hbox{\strut IC4040, D169}\hbox{\strut CGCG 160-252}}& 195.158042, 28.057361 & 65 & 0.02615 & 872 & 100.8  & 335 & 0.45 & early\\

GMP~3779 & \vtop{\hbox{\strut D54, MRK58}\hbox{\strut CGCG 160-073}}& 194.772000, 27.644583 & 29 & 0.01853 & -1316 & 46.8  & 634 & 0.82 & advanced\\

GMP~3071 & D66 & 194.983804, 27.746328 & 71 & 0.02963 & 1860 & 19.1 & 402 & $>$0.35 & advanced\\
\hline  
\end{tabular}
\tablefoot{RA and Dec coordinates and redshift from NED. Inclination from HyperLeda ($i$ = 90$^{\circ}$ corresponds to an edge-on galaxy). Radial velocity w.r.t. the cluster $V_{Coma}$. Estimated stellar mass from MPA-JHU DR7\footnote{\url{https://wwwmpa.mpa-garching.mpg.de/SDSS/DR7/Data/stellarmass.html}} assuming \citet{Kroupa2001} IMF. $d_{\mathrm{Coma}}$ w.r.t. Coma coordinates RA 194.952917 deg, DEC 27.980555 deg. \ion{H}{i} deficiency from \citet{Gavazzi2018}, lower limit on GMP~3071 from \citet{Chen2020}. Stripping phase according to \citet{Yagi2010}. For NGC~4848, stripping phase was determined using the same definition.}

\end{table*}

 \subsection{IRAM 30m observations}  
Observations were carried out with the IRAM 30m antenna operated by Institute de Radio Astronomie Millimétriqie located at Pico Velenta, Spain, between December 2014 and December 2016. Two prominent CO lines: $^{12}$CO(1-0) ($\nu_{rest}$~=~115.271~GHz) and $^{12}$CO(2-1) ($\nu_{rest}$~=~230.538~GHz) were observed simultaneously with a spectral line receiver EMIR in E090 and E230 bands. To EMIR, two different backends were connected; the Fast Fourier Transform Spectrometers (FTS) with a spectral resolution of ~200 kHz; and a wideband autocorrelator WILMA with a spectral resolution of 2 MHz as a backup to prevent potential problems with data platforming between the FTS units, as recommended by IRAM. The Wobbler switching mode, which utilizes a movement of the secondary reflector to switch between ON and OFF-positions, was chosen for the observation. To avoid an accidental targeting of tails of observed galaxies in OFF-positions, the maximum throw of the reflector was used.\par
The half power beam width (HPBW) of the main beam of the IRAM 30m telescope can be determined as HPBW(")~=~2460/$\nu$(GHz), resulting in HPBW $\sim$ 21.7" $\approx$ 10.3 kpc and 10.9" $\approx$ 5.2 kpc for CO(1-0) and CO(2-1), respectively. The main beam projected area corresponds to $\Omega_{B}$ $\approx$ 120 kpc$^2$ and 31 kpc$^2$, respectively, taking the 1/ln2 factor for the Gaussian beam shape correction into account.\par
The list of observed positions with on-source observing times can be found in Table \ref{table_positions}. For each galaxy, we covered the central part of the galaxy and one to three regions in its outer part (except for GMP~3071 where no outer region was observed due to limited observing time). The selection of the integration points generally followed the brightest regions of the H$\alpha$ map.\par 
The data were reduced by engaging CLASS tool of GILDAS software package developed for primary use in IRAM data reduction. Because of the known platforming issue of the FTS backend, we used a script FtsPlatformingCorrection5.class to perform an automatic correction by subtracting baselines from the individual FTS units. However, the observed lines were approximately in the center of our spectra and therefore were not noticeably affected by this issue. All scans have been examined manually and poor-quality ones were excluded from the analysis. The primarily first-order baselines were subtracted from the data utilizing the line-free channels. The original corrected antenna temperature, $T^*_{A}$, was converted to main-beam brightness temperature following $T_{mb}$ = $T^*_{A}F_{eff}/\eta_{mb}$, where the main-beam efficiency was $\eta_{mb}$ = 0.78 at CO(1-0) and 0.59 at CO(2-1) rest frequency and forward efficiency of $F_{eff}$ = 0.94 and 0.92, respectively. The spectra were smoothed to 30.48 km s$^{-1}$ channels. The conversion factor between antenna temperature and flux density is $S_{\nu}/T^*_{A} = 3514/(\eta_{a}D^2)$, where $D$ is the diameter of the telescope and the telescope aperture efficiency of $\eta_{a} \sim$ 0.6 at CO(1-0) and $\sim$ 0.41 at CO(2-1) rest frequency. Using main-beam brightness temperature, the relation becomes $S_{\nu}/T_{mb} = \frac{3514}{\eta_{a}D^2}\frac{\eta_{mb}}{F_{eff}}$. The resulting conversion factor is $\sim$ 5 Jy beam$^{-1}$ K$^{-1}$ for both bands. 
The conversion between CO antenna temperature and flux density depends on the spatial distribution of the CO emission within the IRAM 30 m beam. Our flux density conversion adopts the IRAM 30 m point-source conversion factor, appropriate for compact emission located near the pointing center. Since our observations consist of isolated pointings (rather than fully sampled CO maps), the true source-beam coupling is uncertain, especially in the stripped regions, where the CO emission may be clumpy, partially extended, or offset from the nominal pointing center. If the emission did instead uniformly fill the main beam, the inferred fluxes would be higher by approximately 30$\%$ relative to the point-source conversion. We therefore included an additional systematic uncertainty on the order of 30$\%$ associated with the temperature-to-flux conversion and unknown source-beam coupling. We adopted the larger of the systematic uncertainty and the Gaussian-fit uncertainty.

\begin{table*}
\caption{Observed positions.}             
\label{table_positions}
\centering          
\begin{tabular}{l l c c c c c}
\hline\hline       
Object & \tstrut Region & \vtop{\hbox{\strut RA}\hbox{\strut (J2000)}} & \vtop{\hbox{\strut Dec}\hbox{\strut (J2000)}} & \vtop{\hbox{\strut $d_{\mathrm{parent}}$}\hbox{\strut [kpc]}} & \vtop{\hbox{\strut $t_\mathrm{{CO(1-0)}}$ }\hbox{\strut [hr]}} & \vtop{\hbox{\strut $t_\mathrm{{CO(2-1)}}$ }\hbox{\strut [hr]}}\\ 
\hline

\tstrut NGC~4848  & disk & 12:58:05.60 & 28:14:34.0 & 0 & 2.0 & 2.0\\
& T1 & 12:58:03.61 & 28:14:51.7 & 15.0 & 6.8 & 5.2\\
& T2 & 12:58:03.95 & 28:15:14.8 & 21.9 & 12.4 & 11.1\\
& T3 & 12:58:01.17 & 28:15:42.3 & 42.5 & 13.0 & 10.9 \bstrut\\ \hline
\tstrut GMP~3896  & disk & 12:58:55.96 & 27:50:00.1 & 0 & 1.7 & 1.7\\
& TIN & 12:58:54.68 & 27:49:48.2 & 9.8 & 5.2 & 5.2\\
& T1 & 12:58:53.21 & 27:49:33.7 & 21.3 & 5.4 & 5.4 \bstrut\\ \hline
\tstrut GMP~2559  & disk & 13:00:37.86 & 28:03:29 & 0 & 1.3 & 1.3\\
& SE-F1 & 13:00:38.95 & 28:03:02.5 & 14.3 & 5.4 & 6.0\\
& S-K2 & 13:00:36.55 & 28:02:54.1 & 18.5 & 2.6 & 2.9 \bstrut\\ \hline
\tstrut GMP~3779  & disk & 12:59:05.27 & 27:38:39.7 & 0 & 1.3 & 1.3\\
& T1 & 12:59:03.23 & 27:38:18.5 & 16.3 & 7.3 & 7.3 \bstrut\\ \hline
\tstrut GMP~3071  & disk & 12:59:56.16 & 27:44:47.3 & 0 & 4.5 & 5.4\bstrut\\

\hline  
\end{tabular}
\tablefoot{Coordinates, projected distance from the parent galaxy and on-source time after the exclusion of bad scans of both lines for observed regions.}

\end{table*}

\section{Results}
\label{section_results}
Galaxies NGC~4848, GMP~2559, GMP~3896, and GMP~3779 were detected in both CO(1-0) and CO(2-1) emission. In GMP~3071, we recovered a marginal detection in CO(1-0) only with integrated S/N of 2.9. Moreover, in galaxies NGC~4848, GMP~2559, and GMP~3779, molecular gas was detected clearly outside the galaxy in their ram pressure-stripped tails. CO lines were fitted with Gaussians using the GILDAS CLASS software. Fitted spectra can be found in Fig. \ref{fig_4848_with_spectra}-\ref{fig_3071_with_spectra}. The fit parameters can be found in Table \ref{table_results}. From this, the CO luminosity and molecular gas mass were determined. The H$_2$ mass was calculated from CO(1-0) luminosity using 
\begin{equation}
    M_{H_2}  [\mathrm{M_{\odot}}] = 4.3 L'_{CO}  [\mathrm{K}\mathrm{km}\mathrm{s^{-1}}\mathrm{pc^2}],
\end{equation}
where we used the standard Milky Way CO-to-H$_2$ conversion factor ($\alpha_{CO}$) of 4.3 M$_{\odot}$ (K km s$^{-1}$ pc$^2$)$^{-1}$ \citep{Bolatto2013}. The results based on different conversion factors are discussed in Sect.~\ref{section_expectation}. For CO(2-1), we assumed the typical value for the CO(2–1)-to-CO(1–0) ratio of 0.8 \citep{Leroy2009}. For the CO luminosity $L^{'}_{CO}$, we followed \citep{Solomon2005} the formula,
\begin{equation}
    L'_{CO} = 3.25 \times 10^7 S_{CO} \Delta \upsilon \nu^{-2}_{obs} D^2_L (1+z)^{-3},
\end{equation}
where $S_{CO} \Delta \upsilon$ is the CO velocity integrated line flux in Jy km s$^{-1}$, $v_{obs}$ is the observed CO line frequency in GHz, $D_L$ is the distance in Mpc, and $z$ is the redshift. The quoted CO luminosities and molecular gas masses are based on the compact-source flux conversion; for the beam-filling emission, they would increase by roughly 30$\%$, while compact emission offset from the beam center could require a larger correction.
\par
In this step, two definitions for the signal-to-noise ratio (S/N) are introduced. The peak S/N is defined as 
\begin{equation}
    \mathrm{S/N}_{\mathrm{peak}} = \frac{T_{{\mathrm{peak}}}}{rms},
\end{equation}
whereas the integrated S/N is defined as
\begin{equation}
    \mathrm{S/N}_{\mathrm{int}} = \frac{I_{CO}}{rms\sqrt{FWHM \delta v}},
\end{equation}
where $I_{CO}$ is the integrated intensity and $\delta v$ is smoothed channel width. We used S/N$_{int}$ to quantify the significance of the total integrated CO emission, including weak emission distributed over multiple spectral channels. The root-mean-square (rms) noise was calculated by CLASS from the baseline-subtracted, line-free channels outside the defined CO line windows. For detections with FWHM~$<$~3$\delta v$, we used the peak S/N to characterize the significance of the detection.

\subsection{NGC~4848}
\label{subsection_NGC4848}
In the disk of galaxy NGC~4848, the detected mass of the molecular gas is M$_{H_2, disk}$ = (50 $\pm$ 15) $\times$ 10$^8$ M$_{\odot}$ based on the CO(1-0) luminosity which is comparable with previous CO(1-0) observation by \citet{Vollmer2001} with reported total mass of  M$_{H_2}$~=~42~$\times$~10$^8$~M$_{\odot}$ using the Milky Way conversion factor. The CO(1-0) emission coming from the inner $\sim 5$~kpc of the disk has a distinct double-component (double-horn) profile with a slightly stronger peak on the high-velocity side. The asymmetry is even more pronounced in the CO(2-1) spectrum where the high-velocity part of the central peak clearly dominates its shape. \citet{Vollmer2001} showed that the southern disk side is redshifted, while the northern is blueshifted.\par
\citet{Vollmer2001} further stated that there is a secondary CO peak $\sim$8 kpc NW from the optical center with a mass of M$_{H_2}$~=~3.1~$\times$~10$^8$~M$_{\odot}$ that roughly coincides with \ion{H}{i} peak detected by \citet{BravoAlfaro2001} and also the X-ray extension reported by \citet{Sun2021}. This region only has a marginal cross section with our CO(1-0) beam, so we could not recover this observation, as our main focus was on previously unexplored tail regions.\par

In the tail, we see molecular gas-rich regions not identified in prior studies up to 40--50 kpc from the main body (i.e., T1, T2, and T3) totaling M$_{H_2, tail}$ = (4.8 $\pm$ 1.4) $\times$ 10$^8$ M$_{\odot}$ which is approximately 10~$\%$ of the mass in the disk. In both CO(1-0) and CO(2-1), all detected lines show similar velocities with no clear gradient. This could be explained by a low radial velocity of NGC~4848 relative to the cluster (i.e., the galaxy moves mostly in the plane of the sky). As in the case of the galaxy disk detection, some tail regions also show substructure. While in T1 region, the second peak is only suggested by a number of positive channels right from the primary peak, in T2 it is clearly visible in both lines. Since the velocity difference of the line components is comparable to the galaxy rotational velocity of 270 km s$^{-1}$ \citep{Vollmer2001}, they probably correspond to gas stripped from the opposing disk sides, blending within the IRAM beam. As already stated, the southern disk side is receding and the northern is approaching. The presence of the dust extinction in the optical HST image of NGC~4848 indicates that the western disk side is the near side. Since the ICM wind angle is slightly tilted toward the west relative to the galaxy major axis (as seen in the H$\alpha$ image, Fig. \ref{fig_4848_with_spectra}) and ram pressure is stronger on the side where the wind opposes the rotation of the galaxy, the blueshift of the T1 and T3 detections is expected. In the T2 region, the weaker redshifted detection could trace a gas from the other side of the disk. A total mass of M$_{H_2, tot}$ = (55 $\pm$ 16) $\times$ 10$^8$ M$_{\odot}$ was detected in this galaxy.

\subsection{GMP~3896}
\label{subsection_GMP3896}
In GMP~3896, the disk molecular gas mass of M$_{H_2, disk}$ = (18.5 $\pm$ 5.6) $\times$ 10$^8$ M$_{\odot}$ was detected. The CO lines are wide, corresponding to the near-edge-on orientation of the rotating galaxy. A spectrum from the inner-tail region (TIN) covering the western edge of the disk indicates that the western disk side is receding. The central CO(1-0) spectrum is highly asymmetric and has three distinct components, with the blueshifted component the strongest, suggesting that gas on the leading (blueshifted) side of the galaxy is pushed by ram pressure toward the center and compressed.\par
Outside of the central region, CO was only detected on the boundary between the disk and the tail (TIN region), with M$_{H_2, inner tail}$ = (2.1 $\pm$ 0.6) $\times$ 10$^8$ M$_{\odot}$. The CO(1-0) and CO(2-1) detections show similar features, although in CO(2-1), the inner tail detection is considerably noisy. For the tail region, only an upper limit was measured, corresponding to M$_{H_2, T1}$ = 0.6 $\times$ 10$^8$ M$_{\odot}$ assuming a linewidth of 100 km s$^{-1}$. According to \citet{Yagi2010}, GMP~3896 does not exhibit any disk star formation outside of its central region; however, the H$\alpha$ is extended far from the disk. No molecular gas in this region was recovered by our observation.
Total mass of M$_{H_2, tot}$ = (20.6 $\pm$ 6.2) $\times$ 10$^8$ M$_{\odot}$ was detected in this galaxy with additional 0.6 $\times$ 10$^8$ M$_{\odot}$ as an upper limit.

\subsection{GMP~2559}
\label{subsection_NGC2559}
In GMP~2559, M$_{H_2, disk}$ = (31.3 $\pm$ 9.4) $\times$ 10$^8$ M$_{\odot}$ was detected in the disk. The detection is highly asymmetrical double-horn with a dominant blueshifted part. From the VLA observations by \citet{BravoAlfaro2001}, it is known that the \ion{H}{i} peak is offset from the galaxy center toward the south-east, in the same direction as the H$\alpha$ tail. This suggests that the molecular gas may be pushed/compressed at the leading side of the galaxy in a similar way, which can result in blue feature we see in the spectrum. The overall morphology indicates that more than one environmental mechanism is at play, as RPS alone is unlikely to produce features observed across different wavelengths.\par

The structure of the H$\alpha$ tail is rather complex. The south-eastern region SE-F1, which is named in agreement with \citet{Yoshida2012}, shows a rich substructure in its spectrum consisting of one peak on the parent galaxy velocity and three weaker blueshifted peaks. In \cite{Yoshida2012}, a low-velocity H$\alpha$ component ($\sim$1000~km~s$^{-1}$ with respect to the galaxy center) was revealed, likely associated with a combined action of RPS and tidal interaction, as witnessed by the complex morphology of the H$\alpha$ emission. Searching in the same velocity range, we recovered this feature with a marginal detection with S/N$_{peak}$ of 1.8. The blueshift of the stripped material is expected from the velocity of the galaxy with respect to the cluster. In CO(2-1), we only recovered two of the CO(1-0) detections. There are multiple possible causes: for instance, the structure of the CO gas could be very clumpy and the clumps responsible for the CO(1-0) detections might not be encompassed by a smaller CO(2-1) beam; alternatively, the molecular gas might not be dense or warm enough for CO(2-1) excitation or the CO(2-1) is too faint and is diluted by the beam. In the southern region S-K2, only a low S/N peak in CO(1-0) was detected with a velocity ~150 km s$^{-1}$ lower than the parent galaxy. The total mass of tail molecular gas was M$_{H_2, tail}$ = (6.5 $\pm$ 2.0) $\times$ 10$^8$ M$_{\odot}$, which is approximately 20$\%$ of disk gas mass. A total mass of M$_{H_2, tot}$ = (37.8 $\pm$ 11.4) $\times$ 10$^8$ M$_{\odot}$ was detected in this galaxy.

\subsection{GMP~3779}
\label{subsection_GMP3779}
In the disk of GMP~3779, M$_{H_2, disk}$ = (12.9 $\pm$ 3.9) $\times$ 10$^8$ M$_{\odot}$ was found. In the tail of GMP~3779 at a distance of 17 kpc from the galaxy nucleus, a clear detection at nearly the exact same velocity as the parent galaxy can be seen in both CO(1-0) and CO(2-1). Although GMP 3779 has a large radial velocity (more than -1300 km s$^{-1}$ relative to the cluster), the near-systemic velocity of the CO detection at a projected distance of 17 kpc is not necessarily unexpected. In hydrodynamical simulations of ram-pressure-stripped wakes, stripped gas is accelerated progressively by the ICM wind, with the acceleration completed only at distances on the order of $\sim$100 kpc behind the galaxy; denser ISM clouds are accelerated more slowly than less dense gas \citep{Roediger2008}. The observed CO may therefore trace recently stripped or slowly accelerated dense gas. Another possibility is that we observe a case of a stripped gas cloud that is falling back into its parent galaxy, as previously reported by \citet{Cramer2021} and \citet{Souchereau2025}. The total tail molecular gas mass was M$_{H_2, tail}$ = (1.3 $\pm$ 0.4) $\times$ 10$^8$ M$_{\odot}$. The total mass of M$_{H_2, tot}$ = (14.2 $\pm$ 4.3) $\times$ 10$^8$ M$_{\odot}$ was detected in this galaxy.

\subsection{GMP~3071}
\label{subsection_GMP3071}
For the last galaxy of the sample, GMP~3071, only the parent galaxy was observed, so no conclusions about its tail gas content can be drawn. Even the detection in the galaxy disk is marginal with S/N$_{int}$ of 2.9 and an estimated molecular gas mass of M$_{H_2, disk}$ = (0.4 $\pm$ 0.3) $\times$ 10$^8$ M$_{\odot}$. This is consistent with H$\alpha$ observations that reveal nearly fully gas-depleted galaxy with star formation only in its central ($\sim$1 kpc) core and stream of gas extending outside to the tail \citep{Yagi2010}.

\section{Molecular gas properties}
\label{section_expectation}
\begin{figure}[!ht]
   \centering
   \includegraphics[width=\hsize]{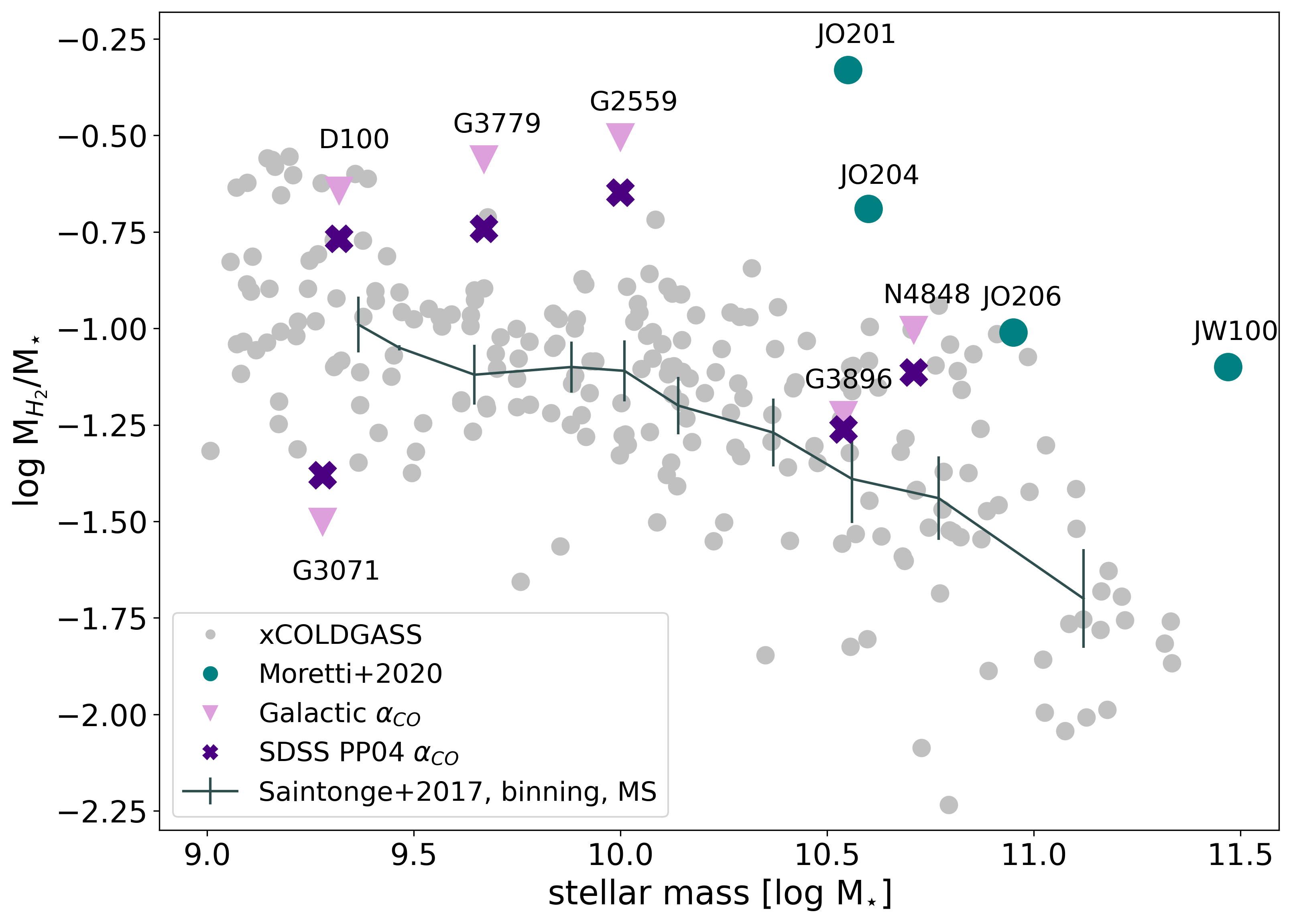}
      \caption{Molecular gas fraction as a function of the stellar mass. The H$_2$ masses from this work are marked by purple crosses and pink triangles using the $\alpha_{CO}$ from \citet{Accurso2017} and the standard Galactic $\alpha_{CO}$, respectively. Data for the galaxy D100 come from \citet{Jachym2017}. The gray line shows the mean scaling relation found in the xCOLDGASS survey by \citet{Saintonge2017} for field galaxies on the main sequence (in accordance with \citealt{Moretti2020}), while the gray dots are the single measurements. Four GASP galaxies from \citet{Moretti2020} are shown in green.}
         \label{fig_molecular_gas_fraction}
   \end{figure}

To better understand the observed molecular gas content in the sample galaxies, we assessed their molecular gas fraction and compare it to that of typical star-forming galaxies unaffected by ram pressure stripping. For the sake of this comparison, only the molecular gas present in the parent galaxy (not in the tail) was taken into account. In Fig. \ref{fig_molecular_gas_fraction}, we show the molecular gas fraction defined as $M_{H_2}$/$M_{\star}$ as a function of the stellar mass. 
 
We compared the molecular gas fraction of our sample of
RPS galaxies with the mean value for star-forming main-sequence galaxies from xCOLDGASS survey \citep{Saintonge2017}, along with the four GASP galaxies with ALMA CO observations, all RPS cluster galaxies
with high stellar masses \citep[M$_\star$ > 3.5 $\times$ 10$^{10}$ M$_{\odot}$;][]{Moretti2020}. As the result is strongly dependent on the CO-to-H${_2}$ conversion factor, we used both the standard Milky Way value of 4.3 $\text{M}_{\odot}$ ($\text{K km s}^{-1} \text{pc}^2)^{-1}$ \citep{Bolatto2013} as well as a metallicity-dependent value following \citet{Accurso2017}. The metallicity of each object was calculated according to \citet{Pettini2004} for which the [OIII], [NII], H$\alpha$, and H$\beta$ emission line fluxes were retrieved from SDSS DR19 \textit{galSpecLine} table \citep{Tremonti2004, Brinchmann2004} and can be found in Table \ref{table_metal}. The choice of conversion factor introduces differences in resulting molecular mass on the order of $\sim30\%$, with variations ranging from $\sim10\%$ to $\sim35\%$.

\begin{table}
\caption{Estimated gas-phase metallicities and adopted CO-to-H$_2$ conversion factors.}
\label{table_metal}
\centering
\begin{tabular}{l c c}
\hline\hline
Name &
$12+\log(\mathrm{O/H})$ &
$\alpha_{\mathrm{CO}}$ \\
&
(PP04) &
[$M_\odot$ (K km s$^{-1}$ pc$^2$)$^{-1}$] \\
\hline
NGC~4848 & $8.772 \pm 0.006$ & 3.34 \\
GMP~3896 & $8.728 \pm 0.020$ & 3.95 \\
GMP~2559 & $8.791 \pm 0.007$ & 3.11 \\
GMP~3779 & $8.817 \pm 0.006$ & 2.83 \\
GMP~3071 & $8.629 \pm 0.018$ & 5.71 \\
D100      & $8.782 \pm 0.007$ & 3.22 \\
\hline
\end{tabular}
\tablefoot{Gas-phase oxygen abundances were estimated using the PP04 O3N2 calibration \citep{Pettini2004}. The CO-to-H$_2$ conversion factors were calculated following \citet{Accurso2017}, adopting the estimated metallicities and the average offset from the star-forming main sequence ($\Delta$MS) for jellyfish galaxies found by \citet{Vulcani2018}.}
\end{table}

\par
In three sample galaxies (NGC~4848, GMP~2559, and GMP~3779), the molecular gas content is higher than expected compared to galaxies of the same stellar mass from a sample of field galaxies and is comparable to the GASP sample. This also applies to galaxy D100 from \citet{Jachym2017}.
For galaxy GMP~3896, we included molecular gas detected in the disk as well as in the border region between the disk and the tail (TIN) because it is impossible to distinguish the exact origin of the gas due to the projection. Even with this caveat, the molecular fraction of this galaxy lies closer to the xCOLDGASS mean value than the previous three cases, while still remaining slightly above. As commented later on, it seems that the total amount of cold gas in this galaxy (H${_2}$ and \ion{H}{i} combined) is lower than expected from field galaxies of the same mass, meaning that some of the cold gas might have been lost in the RPS process. 
GMP~3071, significantly below the mean molecular fraction, is probably in a late stage of stripping and has essentially no molecular gas left. Therefore, the low molecular gas fraction result is not surprising.
\par
The two most massive galaxies in our sample, GMP 3896 and NGC 4848, exhibit H${_2}$-gas fractions more consistent with those of the xCOLDGASS population than do the four GASP galaxies. However, this apparent difference is unlikely to be statistically significant and may reflect differences in the galaxies' orbital properties, stages of RPS, or internal gas processing.\par

The evolution of the molecular component of the ISM experiencing the ram pressure stripping remains unclear. Early observations found that the molecular gas content of Virgo cluster galaxies was relatively normal and not as severely deficient as the \ion{H}{i} \citep{Kenney1989}. More recent observations of \citet{Boselli2014c} and \citet{Fumagalli2009}, among others, revealed a modest H$_2$-deficiency in Virgo galaxies that are severely \ion{H}{i}-deficient. This deficiency was in both cases associated with strong \ion{H}{i}-depletion inside stellar disks of the galaxies with Def$_{HI}$ > 1. 
A modest H$_2$ deficiency in strongly \ion{H}{i}-deficient Virgo galaxies might be expected since stripping acts from the outside-in, and the H$_2$/\ion{H}{i} ratio generally decreases strongly with radius. 
Moreover, high-resolution studies of Virgo cluster galaxies undergoing stripping show that little to no molecular gas remains in the disk once the \ion{H}{i} is fully stripped. In these galaxies, stripping proceeds by removing the lower density gas, including \ion{H}{i} and diffuse molecular gas, leaving behind some denser molecular clouds \citep{Abramson2014, Abramson2016, Cramer2020}. The dense clouds left behind are short-lived (since few have been observed thus far) and only a small fraction of the original gas mass at that radius is found within them. This means that all the ISM including molecular gas is effectively stripped and joins the tail (and star formation is quenched) soon after the lower density gas is removed, even if molecular clouds are not directly stripped initially.
As the RPS process deprives the galaxy of the gaseous halo as well, there is no way for the galaxy to gain new gas from the CGM. For thorough description of this topic, see \citet{Cortese2021}.\par
At the same time, our results and results of the GASP Project \citep{Moretti2020}, among others, show an increased molecular gas fraction compared to the sample of field galaxies. Recently, \citet{Roberts2022} showed the enhancement of the star formation on the leading half of the galaxies (the galaxy side opposite to the direction of the RPS tail) on a sample of 29 jellyfish galaxies in nearby clusters; concluding that it is a result of a compression of atomic gas by the ram pressure leading to the molecular gas formation that, in turn, fuels star formation. Similarly, \citet{Moretti2026} showed that the compression on the leading side produces an efficient conversion of H$_2$ into stars using the resolved depletion time maps.
This conclusion is supported by several other observations (e.g., \citealp{Kenney2004, Roberts2020, Cramer2021}) as well as simulations \citep{Troncoso-Iribarren2020}.\par

Observations suggest that properties of the cold gas component of cluster member galaxies could be cluster-dependent as the halo mass, galaxy number density, and velocity dispersion (along with other effects) create a unique environment in each cluster. The strength of the ram pressure depends on cluster mass both via velocity of the infalling galaxies and density of the ICM. Indeed, there are significant observational differences between gas properties of various cluster members. In the Fornax cluster, galaxies have been reported to be H$_2$-deficient in both cases of undisturbed CO emission and disturbed CO emission (-0.5 dex, -1.1 dex respectively; compared to sample of field galaxies; \citealt{Zabel2019}). In the Virgo cluster (i.e., about ten times more massive) the molecular fraction of member galaxies is normal-to-high compared to the sample of field galaxies \citep{Brown2021}; whereas the \ion{H}{i}-deficiency is similar in both clusters \citep{Loni2021}. However, the most \ion{H}{i}-deficient Virgo galaxies (Def$_{\mathrm{\ion{H}{i}}}$>1) have (on average) a factor of $\sim$2 less H$_2$ gas than field galaxies of the same stellar mass \citep{Boselli2014c}. The Coma cluster is again about ten times more massive than Virgo cluster, so the effects of ram pressure whose strength scales approximately linearly with cluster mass is expected to be even more pronounced; including the RPS-induced gas compression that can lead to the elevated molecular fraction.\par

\begin{figure}[!ht]
   \centering
   \includegraphics[width=\hsize]{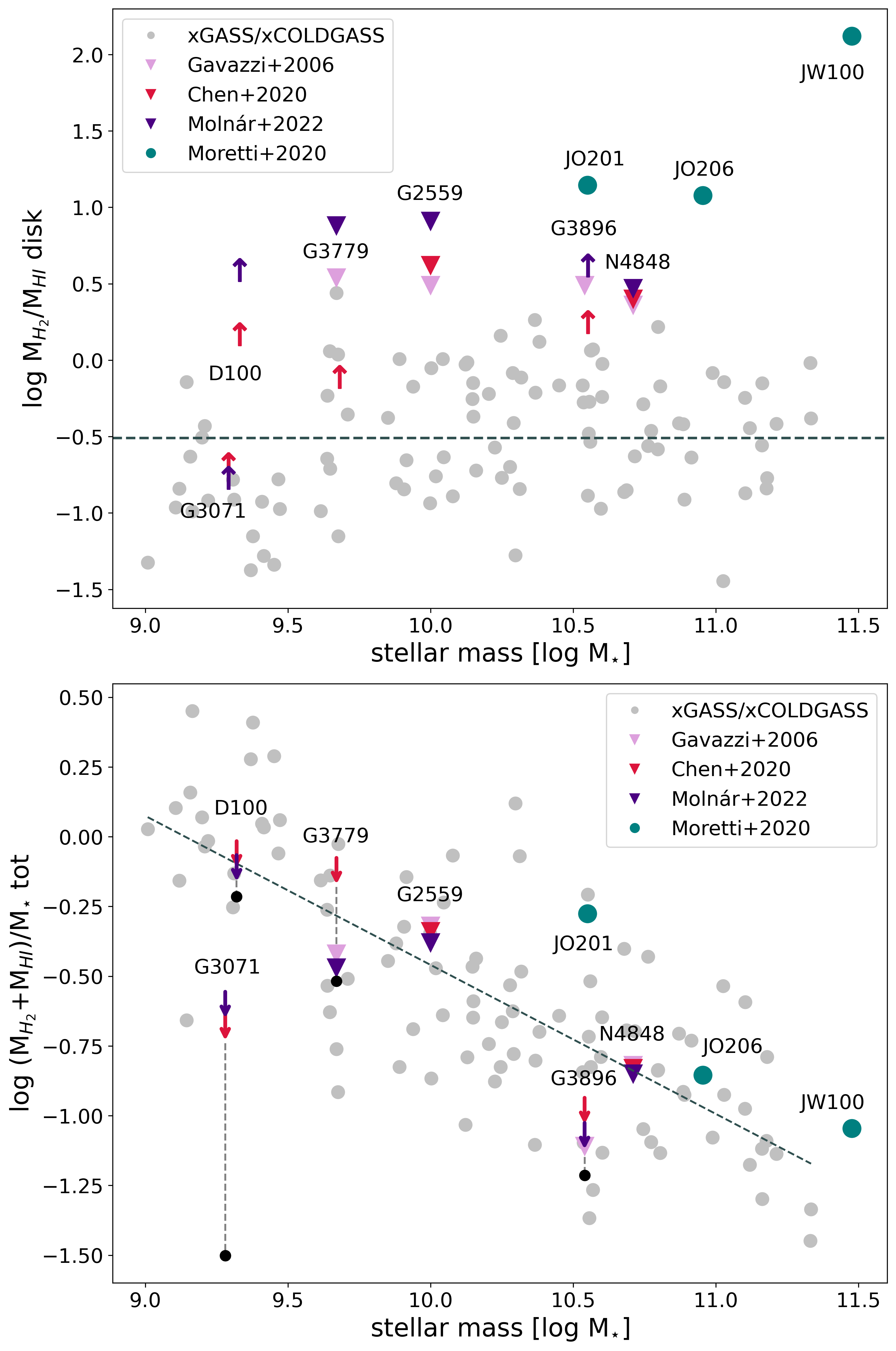}
      \caption{\textit{Top:} Molecular-to-atomic-gas-mass fraction in the disk (M$_{H_2, disk}$/M$_{HI, disk}$) as a function of the stellar mass. \textit{Bottom:} Total (disk+tail) gas fraction ((M$_{H_2, tot}$+M$_{HI, tot}$)/M$_\star$) as a function of the stellar mass. xGASS/xCOLDGASS sample of main-sequence field galaxies is in gray. GASP galaxies from \cite{Moretti2020} are shown in green. Values utilizing \ion{H}{i} observation by \citet{Molnar2022}, \citet{Gavazzi2006}, and \citet{Chen2020} are marked by purple, pink, and red, respectively. The upper and lower limits are marked by arrows. In bottom figure, a range is provided for galaxies with only M$_{HI}$ upper limits available. The lower value is given by M$_{H_2}$ only, whereas the upper value is given by $M_{H_2}$ + respective upper limit on M$_{HI}$.}
         \label{figure_molecular_gas_fraction_II}
   \end{figure}

Another important part of the discussion is a comparison with the atomic gas phase. In Fig. \ref{figure_molecular_gas_fraction_II}, we show the disk molecular-to-atomic gas fraction (M$_{H_2}$/M$_{HI}$) and the total (disk+tail) gas fraction (M$_{H_2, tot}$ + M$_{HI, tot}$)/M$_\star$)  in comparison with a xGASS/xCOLDGASS sample of field main-sequence galaxies detected both in \ion{H}{i} and H$_2$ \citep{Catinella2018}. For galaxies GMP~3071 and D100, we only have upper limits on \ion{H}{i}, resulting in lower limits on M$_{H_2}$/M$_{HI}$ and upper limit on the total gas fraction that are marked with arrows. We used three different \ion{H}{i} observations from \citet{Molnar2022}, \citet{Gavazzi2006}, and \citet{Chen2020} that are marked in purple, pink, and red, respectively. From \citet{Molnar2022}, we estimated upper limits on the \ion{H}{i} mass based on the noise in the Westerbork Synthesis Radio Telescope (WSRT) data cube, assuming spatially unresolved source and velocity dispersion from their absolute i-band magnitude (from SDSS) and the Tully-Fisher relation \citep{Masters2006}. From the upper plot, it can be clearly seen that the amount of the molecular gas compared to the atomic gas exceeds the mean value in all galaxies of our sample; the only exception is the galaxy GMP~3071, which is presumably in a late stage of stripping. What we are likely seeing is the combined effect of more efficient stripping of the atomic phase and an atomic-to-molecular gas conversion that is being enhanced by the ram pressure. 
In the bottom plot, we see the total amount of gas in our sample, including the mass of the molecular gas in tail. For most of the galaxies, the total gas content is normal as compared to field galaxies of the same stellar mass. In this case, the two processes of \ion{H}{i} gas loss are probably compensated for - the loss of \ion{H}{i} due to gas conversion results in gain in H$_2$ mass while the loss of \ion{H}{i} due to stripping can result in increased tail gas mass regardless of the process of tail gas creation. This is supported by the fact that galaxies with the normal gas content (NGC~4848, GMP~2559, and GMP~3779) have CO tail detection, while GMP~3896 only has detection within stellar disk, meaning some of the gas was lost and not recovered. 

\section{Galaxies' evolution within the Coma cluster}
\label{section_history}
In accordance with \citet{Yagi2010}, the sample was designed to follow multiple phases of galaxy evolution within the cluster. Three galaxies show signs of early-phase stripping (NGC~4848, GMP~2559, and GMP~3896) and two find themselves in the advanced-phase stripping (GMP~3779 and GMP~3071). The late stage of stripping as described in \citet{Yagi2010} is not represented in our sample. This is in agreement with projected tail angles of our sample galaxies which are radially outward or nearly azimuthal (Fig. \ref{fig_coma_our_galaxies}), indicating that galaxies are falling inward (NGC 4848, GMP 3896, GMP 3779, GMP 3071) or near pericenter (GMP 2559, D100). That means that our sample is near peak ram pressure. It is of interest to study the molecular gas content of galaxies in different stages to follow the whole process of stripping. However, it is important to stress that the stripping progression of the respective galaxy depends on multiple characteristics such as its mass, velocity, disk-wind angle, and position within the cluster. That makes the categorization a complex issue because two galaxies at about the same orbital position can be at very different stages of gas stripping depending on these characteristics. Galaxies in the advanced-phase group share a low mass and a high radial velocity. The early-phase group is much more diverse in these characteristics. While NGC~4848 is expected to be in an early stage of stripping because of its high mass, low radial velocity and more remote position from the cluster core ($R_{proj}$ $\sim$0.4 $R_{200}$), this is not the case for the rest of the group. GMP~3896 is approximately at the same projected distance as GMP~3071 from the advanced stripping group; however, its mass is $\sim$18 times higher, which could be the main driver of their stripping-phase discrepancy. GMP~2559 exhibits one of the lowest projected distances from the cluster core ($R_{proj}$ $\sim$0.2 $R_{200}$) and (compared to the rest of the early-phase group) it has a relatively low mass. Nonetheless, H$\alpha$ signs of early-phase stripping suggest that its real distance from the cluster core could be greater than the projected estimation.

\subsection{Early-phase stripping group}
Galaxies NGC~4848 and GMP~2559 show an abundance of CO gas in the H$\alpha$-bright tail regions, even though in some of them (i.e., the T3 region of NGC~4848 and the S-K2 region of GMP~2559), the H$\alpha$ features are much more prominent. The last galaxy of the group, GMP~3896, on the other hand, revealed no molecular gas in its tail indicating that it finds itself in a slightly later stage of stripping than the other two galaxies within the same group. Also, comparing these three galaxies in terms of the total gas fraction, GMP~3896 shows greater gas depletion, which again indicates later stripping phase.\par
The history of the galaxy NGC~4848 is not generally agreed upon in the literature. It has a prominent, molecular gas-rich tail despite its relatively large distance from the cluster center, where the efficiency of the ram pressure is not yet high. The assumption that the galaxy NGC~4848 belongs to the early-phase stripping group is supported by the inferred time spent in Coma cluster that would account for its estimated gas depletion. \citet{Fossati2012} using a simulation of \citet{Kapferer2009} argued that the time needed to partially deplete the \ion{H}{i} gas from the galaxy is about 200 Myr, which is shorter than the crossing time in the Coma cluster, supporting the scenario of the first passage through the cluster core. However, \citet{Vollmer2001} posited that the galaxy had entered the cluster $\sim$1 Gyr ago and passed near the cluster center $\sim$400 Myr ago, based on a sticky particles numerical simulation, which included the ram pressure stripping process. Based on the normal total gas fraction (Fig. \ref{figure_molecular_gas_fraction_II}) and the observed gas-rich tail, we lean to the former scenario. Moreover, in \citet{Infurna2026}, it was recently reported that the outskirts of NGC 4848 were quenched $\sim$500 Myr ago, consistent with the first crossing of the galaxy through the cluster center. 
\citet{Vollmer2001} further reported that while the \ion{H}{i} gas was completely swept from the front side of the galaxy, the brightest HII regions are located there. As for the CO gas, it peaks at the center of the disk as expected; however, there is a secondary CO peak shifted from the center toward the tail. The mass of the molecular gas in this region was estimated as $\sim$ 3.3 $\times$ 10$^{8}$ M$_{\odot}$ which is comparable to the mass that we detected in the T1 region of this galaxy. The total molecular gas mass detected by \citet{Vollmer2001} is 4.5 $\times$ 10$^{9}$ M$_{\odot}$.\par
The observations of galaxy GMP~2559 suggest its complex history within the Coma cluster.
The H$\alpha$ tail reveals a two-peak feature with a low-velocity component (-1300 km s$^{-1}$ w.r.t. galaxy velocity) and a fast component (300 km s$^{-1}$ w.r.t. galaxy velocity), both located at south-eastern filament of the tail \citep{Yoshida2012}. Double peak detection is present also in \ion{H}{i} \citet{BravoAlfaro2001} and our CO(1-0) observation. The disturbed optical morphology of GMP 2559 was also identified as consistent with RPS by \citet{Roberts2020}. It is possible that this galaxy has undergone a tidal interaction that caused its complex structure and kinematics in addition to the ongoing ram pressure stripping.

 \subsection{Advanced-phase stripping group}
 GMP~3779 shows at least one region in the tail that is rich in molecular gas and a strong detection in the galaxy disk. However, its \ion{H}{i}-deficiency is among the highest in the sample. While \citet{BravoAlfaro2001} reported an extension of \ion{H}{i} gas density in the stripping direction covering the T1 tail region, the newest \ion{H}{i} study of \citet{Molnar2022} did not show any \ion{H}{i} asymmetry at all. The second galaxy in the advanced-stripping group, GMP~3071, is characterized by a lack of molecular gas in the disk and much smaller stripping radius than GMP~3779. This suggests that this galaxy is probably in the latest phase of stripping from our sample.\par
  We stress the significant difference in the stellar mass of galaxies in the two categories is likely to be the main driver behind their different stripping stages. Even though similar stripping characteristics, such as enhanced H$_2$ content were observed in the advanced-stripping group, RPS presumably acts more abruptly than in the case of more massive galaxies where the gradual outside-in stripping progression can be seen, suggesting that massive galaxies can hold-on to their gas reservoirs for longer \citet{Cortese2021}. 
 
\section{H\texorpdfstring{$\alpha$}{alpha}-CO correlation}
\label{section_halpha}
\begin{figure}[!ht]
   \centering
   \includegraphics[width=\hsize]{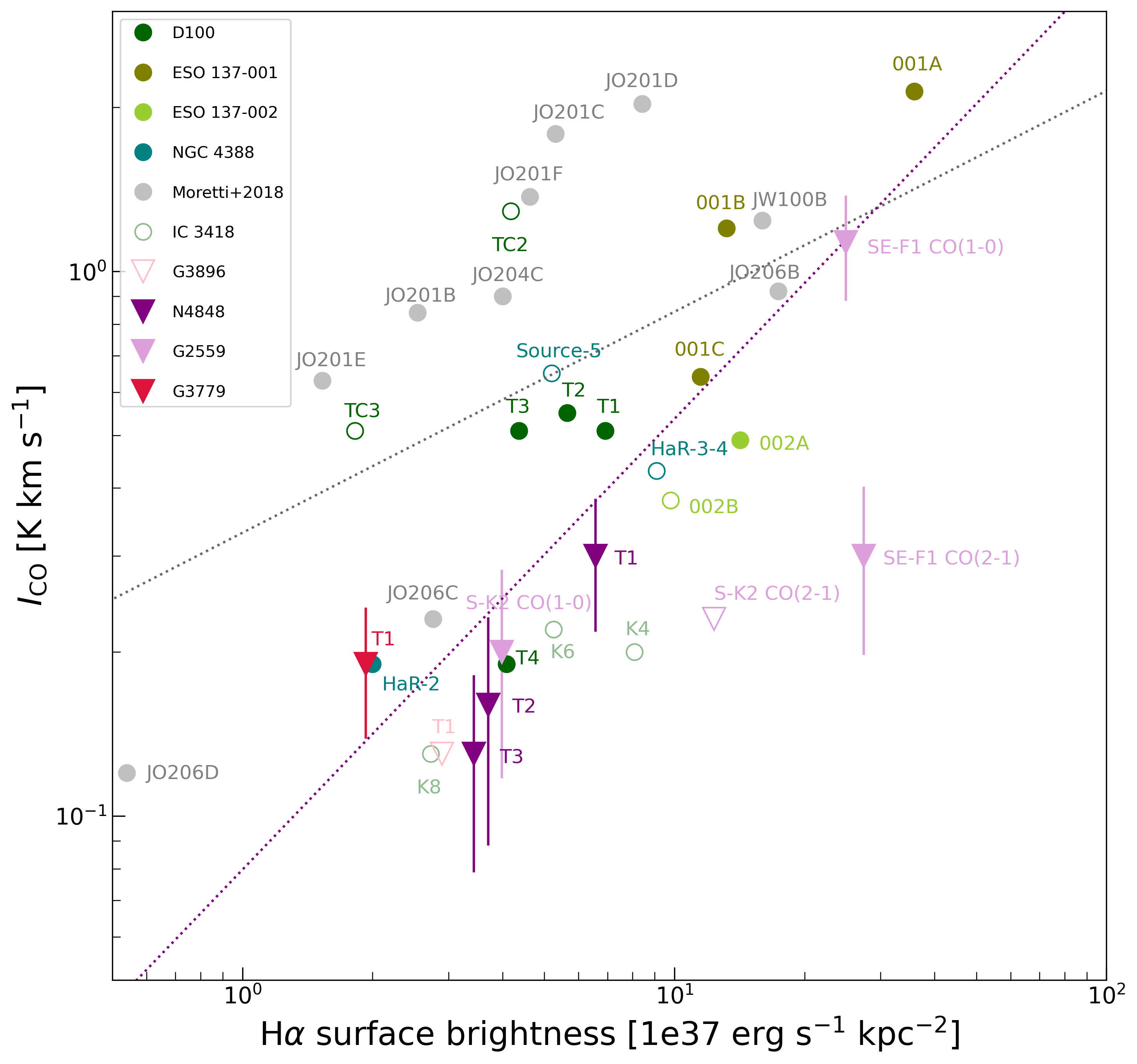}
      \caption{Correlation between CO and H$\alpha$ emission in tails of stripped galaxies. CO(2-1) integrated flux vs. H$\alpha$ surface brightness measured in IRAM 30m apertures (HPBW  $\sim$ 5.2 kpc) from this work are marked by purple-scale triangles. Previously published data from \citet{Jachym2017} and \citet{Moretti2018} are shown as green-scale and gray circles, respectively. Best fit of the data is shown in purple dashed line. Black line shows the fit including the data from \citet{Moretti2018}. For galaxy GMP~2559, CO(1-0) data are shown alongside CO(2-1) as they have significantly better quality. CO upper limits are marked with open symbols.}
         \label{fig_HalphavsCO}
   \end{figure}

 Another useful insight into the evolution of gas of ram-pressure-stripped galaxies is provided by comparison of different gas phases. In Fig. \ref{fig_HalphavsCO}, we compare the beam-weighted CO(2-1) integrated flux and H$\alpha$ surface brightness. The plot features Norma cluster galaxies ESO 137-001 and ESO 137-002, Virgo cluster galaxies, NGC 4388 and IC 3418, and galaxy D100, which shares the Coma cluster with our sample from \citet{Jachym2017}. We also included results from \citet{Moretti2018} for four ram-pressure-stripped galaxies JO201, JO204, JO206, and JW100. The green-scale markers show observations from literature while the pink-scale markers follow this work. Upper limits are plotted using open symbols. For the galaxy GMP~2559, both the CO(1-0) and CO(2-1) flux was used. While CO(2-1) was used for the sake of consistency, CO(1-0) has much higher quality and was therefore more reliable in this context.\par
 The scatter of the plot is more significant than in previous works of \citet{Jachym2017} and \citet{Moretti2018}. One of the reasons is that each author has a different approach to evaluate the H$\alpha$ flux within the beam of the respective radiotelescope. While \citet{Jachym2017} compensated for the tail width, which does not fill the whole IRAM 30m beam, \citet{Moretti2018} only used MUSE spaxels flagged as star-forming or composite within their APEX beams. In this work, we did not include any such compensation and we show H$\alpha$ brightness within the whole IRAM 30m beam. In our case, some of the tail regions are very complex and any estimation of the tail width in a given location would be unreliable. Another reason is related to the previous one: in \citet{Moretti2018}, they only compared both gas phases in regions that are star-forming by using the MUSE spaxel flags. They even went a step further and identified individual star-forming knots within their APEX beams and calculated the H$\alpha$ fluxes within them. In this work, we did not exclude the non-star-forming part of the gas and so, different mechanisms of the gas ionization might be present. Last but not least, the plot also compares different spatial scales: the beam HPBW ranges from $\sim$ 0.9 kpc for Virgo galaxy NGC 4388 to $\sim$ 30 kpc for JO206.\par
 In the previous section, the description of CO detections within H$\alpha$-bright regions was given. Clearly, the gas phases occupy the same space at least at some spatial scales, which is expected as the molecular gas is a prerequisite of the star formation traced by the H$\alpha$. This certainly holds for the compact component of both gas phases, which is linked to the HII regions in the tail \citep{Moretti2026}. However, the amount of HII regions in tails differs significantly from galaxy to galaxy$-$from none (revealed) in ESO 137-002 to possibly a few in galaxies from this work and up to more than 30 in ESO 137-001 \citep{Waldron2023}. From this fact, a question arises as to whether the H$\alpha$-CO correlation still holds outside the HII regions or whether it is even expected to. From Fig. \ref{fig_HalphavsCO}, we see indication that the slope of the correlation might differ for dominantly star-forming gas (gray line) and star-forming and non-star-forming gas combined (purple line).   
 From \citet{Tomicic2021}, we know that the fraction of diffuse ionized gas in galaxy disks is significant (20 -- 80$\%$). For tails, the fraction can reach $\sim$80\% \citep{Tomicic2021b} and it is anti-correlated with tail SFR. This means that the origin of this ionized gas is not necessarily tied to the star formation and can be connected to different ionization mechanisms appearing during the gas stripping and mixing, such as shocks or heating via the ICM. This can be directly studied using different BPT diagnostic diagrams \citep{Poggianti2018, Tomicic2021, Tomicic2021b, Foster2026}. Specifically, the BPT diagram of one of our sample galaxies, GMP 2559, was presented in \citet{Yoshida2012}, with the conclusion that shock-heating does indeed play an important role in ionization of its extended filaments, such as SE-F1 region studied in this work.
 Molecular gas also has a compact and diffuse component. It is known that star formation efficiency (SFE) of RPS tail is usually low \citep{Jachym2014, Moretti2018} and it is not clear what prevents the diffuse part of the cold gas from forming compact regions and igniting star formation. 
 Last but not least, on small scales, the scatter between CO and H$\alpha$ is expected due to different stages of star formation process. While \citet{Moretti2026} recently showed that SFE in H$\alpha$-bright complexes in tail regions follows general Kenniccut-Schmidt relation, it is also possible to observe pre-star forming clouds without any HII regions yet formed as well as prominent HII regions where most of the molecular gas was already used up or dispersed in the later stages of the star formation.
 Therefore, it is important to further study this correlation with high-resolution data from both gas phases which will help discriminate between their different components and search for the actual scales at which their correlation arises, holds, and breaks down.
 
\section{Conclusion}
\label{section_conclusion}

We present new IRAM 30\,m CO(1-0) and CO(2-1) observations of five Coma cluster galaxies undergoing ram pressure stripping, targeting both their disks and extraplanar regions traced by H$\alpha$ emission. Molecular gas was detected in the disks of four galaxies and in the stripped tails of three of them. The molecular gas masses in the tails correspond to approximately 10--20\% of the molecular gas content of the disks.\par
The detection of significant amounts of molecular gas far from the stellar disks provides direct observational evidence that molecular gas can survive ram pressure stripping and/or form in situ within the stripped material. Together with previous detections in Coma, Virgo, Norma, and other clusters, our results demonstrate that molecular gas-rich tails are not rare exceptions but a common outcome of strong ram pressure events.\par
When compared to field scaling relations, several galaxies in our sample show elevated molecular gas fractions at fixed stellar mass. This supports a scenario in which ram pressure not only removes gas, but also compresses the remaining atomic gas in the disks, enhancing its conversion into molecular form. At the same time, the molecular-to-atomic gas ratios are systematically higher than in field galaxies, while the total gas fractions are often close to normal once the gas in the tails is included. This indicates that the apparent molecular gas enhancement arises from a combination of efficient \ion{H}{i} stripping and ram pressure-induced phase transformation, rather than from an overall increase in the total gas reservoir.\par
By comparing CO and H$\alpha$ emission in the stripped regions, we find that molecular gas is preferentially associated with H$\alpha$-bright structures, but with a significantly larger scatter than reported in earlier studies. This likely reflects the coexistence of compact, star-forming molecular gas and a more diffuse molecular component with low star formation efficiency in the tails. Disentangling these components and understanding the physical mechanisms that regulate star formation in stripped molecular gas will require higher-resolution observations.\par
Overall, our study emphasizes the complex impact of ram pressure stripping on the cold ISM of cluster galaxies. While ram pressure efficiently removes gas and contributes to long-term quenching, it can simultaneously stimulate molecular gas formation both within galaxy disks and in the stripped material, temporarily sustaining star formation in otherwise hostile environments. Future spatially resolved CO and multiwavelength observations will be essential to fully characterize the lifecycle of molecular gas during ram pressure stripping and its role in galaxy evolution in dense environments.

\begin{acknowledgements}
A. Srbljanović acknowledges support from the project LM2023059 of the Ministry of Education, Youth, and Sports of the Czech Republic.
P. Jáchym and A. Shaji acknowledge support from the project 25-19512L of the Czech Science Foundation and the institutional project RVO:67985815. We also acknowledge the support by the NSF grant 2407821.
B. Deshev acknowledges support by the Estonian Research Council (grant PSG1200).
This work is based on observations carried out under project numbers 195-14 and 173-16 with the IRAM 30-meter telescope. IRAM is supported by INSU/CNRS (France), MPG (Germany) and IGN (Spain).
This research is based in part on data collected at the Subaru Telescope, which is operated by the National Astronomical Observatory of Japan. We are honored and grateful for the opportunity of observing the Universe from Maunakea, which has the cultural, historical, and natural significance in Hawaii.
This research has made use of the NASA/IPAC Extragalactic Database (NED) which is operated by the Jet Propulsion Laboratory, California Institute of Technology, under contract with the National Aeronautics and Space Administration. We further acknowledge the usage of the HyperLeda database (http://leda.univ-lyon1.fr). \textit{Facility:} IRAM 30m.
     
\end{acknowledgements}

\bibliographystyle{aa}
\bibliography{bib}

\begin{appendix}

\section{CO detections}
\label{section_co_detections}

\begin{table*}
\caption{Results.}
\label{table_results}
\centering
\begin{tabular}{l l c c c c c c c c}
\hline\hline
\hline
\tstrut Source & Line &
\vtop{\hbox{\strut rms}\hbox{\strut [mK]}} &
\vtop{\hbox{\strut Velocity}\hbox{\strut [km s$^{-1}$]}} &
\vtop{\hbox{\strut FWHM}\hbox{\strut [km s$^{-1}$]}} &
\vtop{\hbox{\strut $T_\mathrm{peak}$}\hbox{\strut [mK]}} &
\vtop{\hbox{\strut S/N}\hbox{\strut peak}} &
\vtop{\hbox{\strut S/N}\hbox{\strut int.}} &
\vtop{\hbox{\strut $L'_{\scriptscriptstyle \mathrm{CO}}$}\hbox{\strut [10$^7$ K km s$^{-1}$ pc$^{2}$]}} &
\vtop{\hbox{\strut $M_\mathrm{H_2}$}\hbox{\strut [10$^8$ M$_{\odot}$]\bstrut}} \\
\hline
\hline

NGC~4848 & \tstrut CO(1-0) & 1.4 & 7048 $\pm$ 4 & 378 $\pm$ 7 & 22.1 & 15.8 & 65.2 & 117 $\pm$ 35 & 50 $\pm$ 15\\
& CO(2-1) & 2.4 & 7052 $\pm$ 2 & 327 $\pm$ 5 & 47.9 & 20.0 & 69.3 & 50 $\pm$ 15 & 26.7 $\pm$ 8.0\\
\hline

T1 & \tstrut CO(1-0) & 0.4 & 6874 $\pm$ 6 & 110 $\pm$ 12 & 4.2 & 10.5 & 21.2 & 5.9 $\pm$ 1.8 & 2.5 $\pm$ 0.8\\
& CO(2-1) & 0.9 & 6853 $\pm$ 21 & 148 $\pm$ 46 & 1.9 & 2.1 & 4.8 & 0.9 $\pm$ 0.3 & 0.5 $\pm$ 0.1\\
\hline

T2 & \tstrut CO(1-0) & 0.4 & 6859 $\pm$ 6 & 74 $\pm$ 15 & 2.0 & 5.0 & 8.9 & 2.0 $\pm$ 0.6 & 0.9 $\pm$ 0.3\\
& & 0.4 & 7082 $\pm$ 22 & 171 $\pm$ 54 & 0.8 & 2.0 & 5.2 & 1.8 $\pm$ 0.5 & 0.8 $\pm$ 0.2\\
& CO(2-1) & 0.6 & 6829 $\pm$ 5 & 30 $\pm$ 60 & 2.4 & 4.0 & 4.4 & 0.2 $\pm$ 0.1 & 0.1 $\pm$ 0.1\\
& & 0.6 & 7082 $\pm$ 12 & 57 $\pm$ 36 & 1.3 & 2.2 & 3.2 & 0.2 $\pm$ 0.1 & 0.1 $\pm$ 0.1\\
\hline

T3 & \tstrut CO(1-0) & 0.3 & 6978 $\pm$ 22 & 134 $\pm$ 40 & 0.8 & 2.7 & 6.3 & 1.4 $\pm$ 0.4 & 0.6 $\pm$ 0.2\\
& CO(2-1) & 0.5 & 7001 $\pm$ 28 & 139 $\pm$ 61 & 0.9 & 1.8 & 4.0 & 0.4 $\pm$ 0.1 & 0.2 $\pm$ 0.1\\

\hline
\hline

GMP~3896 & \tstrut CO(1-0) & 1.9 & 7277 $\pm$ 9 & 350 $\pm$ 16 & 12.3 & 6.5 & 18.3 & 43 $\pm$ 13 & 18.5 $\pm$ 5.6\\
& CO(2-1) & 4.7 & 7328 $\pm$ 10 & 349 $\pm$ 19 & 19.6 & 4.2 & 15.3 & 22.1 $\pm$ 6.6 & 11.9 $\pm$ 3.6\\
\hline

TIN & \tstrut CO(1-0) & 0.6 & 7465 $\pm$ 11 & 138 $\pm$ 22 & 2.8 & 4.7 & 10.3 & 4.8 $\pm$ 1.4 & 2.1 $\pm$ 0.6\\
& CO(2-1) & 2.1 & 7477 $\pm$ 23 & 74 $\pm$ 43 & 3.8 & 1.8 & 3.0 & 0.9 $\pm$ 0.3 & 0.5 $\pm$ 0.2\\
\hline

T1 & \tstrut CO(1-0)* & 0.7 & ... & ... & ... & ... & ... & 1.3 & 0.6\\
& CO(2-1)* & 0.8 & ... & ... & ... & ... & ... & 1.5 & 0.8\\

\hline
\hline

GMP~2559 & \tstrut CO(1-0) & 1.4 & 7536 $\pm$ 3 & 234 $\pm$ 6 & 27.2 & 19.4 & 51.6 & 73 $\pm$ 22 & 31.3 $\pm$ 9.4\\
& CO(2-1) & 3.6 & 7498 $\pm$ 2 & 178 $\pm$ 5 & 53.6 & 14.9 & 38.1 & 30.1 $\pm$ 9.0 & 16.2 $\pm$ 4.9\\
\hline

SE-F1 & \tstrut CO(1-0) & 0.6 & 6468 $\pm$ 22 & 72 $\pm$ 36 & 1.1 & 1.8 & 2.8 & 1.0 $\pm$ 0.5 & 0.4 $\pm$ 0.2\\
& & 0.6 & 7011 $\pm$ 12 & 58 $\pm$ 17 & 1.6 & 2.7 & 3.6 & 1.1 $\pm$ 0.4 & 0.5 $\pm$ 0.2\\
& & 0.6 & 7239 $\pm$ 18 & 174 $\pm$ 51 & 1.7 & 2.8 & 7.1 & 3.7 $\pm$ 1.1 & 1.6 $\pm$ 0.5\\
& & 0.6 & 7606 $\pm$ 19 & 272 $\pm$ 50 & 2.1 & 3.5 & 11.2 & 7.3 $\pm$ 2.2 & 3.1 $\pm$ 0.9\\
& CO(2-1) & 0.7 & 7002 $\pm$ 13 & 72 $\pm$ 22 & 1.9 & 2.7 & 4.6 & 0.4 $\pm$ 0.1 & 0.2 $\pm$ 0.1\\
& & 0.7 & 7182 $\pm$ 19 & 93 $\pm$ 31 & 1.4 & 2.0 & 3.8 & 0.4 $\pm$ 0.1 & 0.2 $\pm$ 0.1\\
\hline

S-K2 & \tstrut CO(1-0) & 0.8 & 7387 $\pm$ 17 & 75 $\pm$ 50 & 2.3 & 2.9 & 5.0 & 2.3 $\pm$ 1.0 & 1.0 $\pm$ 0.4\\
& CO(2-1)* & 1.3 & ... & ... & ... & ... & ... & 2.6 & 1.4\\

\hline
\hline

GMP~3779 & \tstrut CO(1-0) & 1.4 & 5340 $\pm$ 2 & 86 $\pm$ 4 & 26.8 & 19.1 & 34.9 & 30.0 $\pm$ 9.0 & 12.9 $\pm$ 3.9\\
& CO(2-1) & 3.2 & 5336 $\pm$ 1 & 74 $\pm$ 3 & 48.2 & 15.1 & 25.0 & 11.4 $\pm$ 3.4 & 6.1 $\pm$ 1.8\\
\hline

T1 & \tstrut CO(1-0) & 0.5 & 5334 $\pm$ 11 & 118 $\pm$ 43 & 2.0 & 4.0 & 8.3 & 3.0 $\pm$ 0.9 & 1.3 $\pm$ 0.4\\
& CO(2-1) & 1.0 & 5362 $\pm$ 14 & 87 $\pm$ 22 & 2.1 & 2.1 & 3.7 & 0.6 $\pm$ 0.2 & 0.3 $\pm$ 0.1\\

\hline
\hline

GMP~3071 & \tstrut CO(1-0) & 0.4 & 8618 $\pm$ 40 & 121 $\pm$ 96 & 0.58 & 1.5 & 2.9 & 0.8 $\pm$ 0.6 & 0.4 $\pm$ 0.3\\
& CO(2-1)* & 1.2 & ... & ... & ... & ... & ... & 2.4 & 1.3\\

\hline
\hline
\end{tabular}

\vspace{2mm}
\tablefoot{Total (disk+tail) H$_2$ masses can be found in Sects.
\ref{subsection_NGC4848}--\ref{subsection_GMP3071}. Non-detections with upper limits on
$L'_{\scriptscriptstyle \mathrm{CO}}$ and $M_\mathrm{H_2}$ are marked with *. S/N$_{peak}$ and S/N$_{int}$ defined at the beginning of this section.}
\end{table*}

\begin{minipage}{\textwidth}
  \centering
  \includegraphics[width=0.88\textwidth]{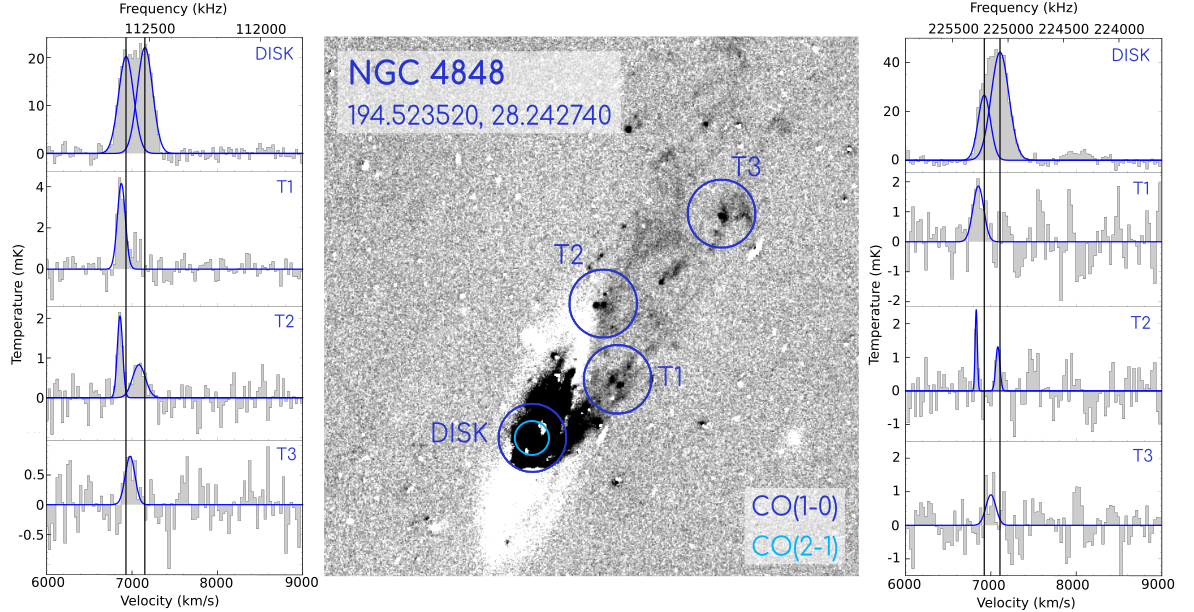}
  \captionof{figure}{Galaxy NGC~4848. Middle: Optical H$\alpha$ image of the galaxy NGC~4848 (Subaru telescope) overlaid with $^{12}$CO(1-0) beams (HPBW $\approx$ 10.3 kpc) of the IRAM 30m antenna located at the observed pointings at the main body of the galaxy and the tail (dark blue). $^{12}$CO~(2-1) beam (HPBW $\approx$ 5.2 kpc) is shown on the central pointing for each galaxy (light blue). $^{12}$CO(1-0) spectra (left) and $^{12}$CO(2-1) spectra (right) smoothed to 30.48 km s$^{-1}$ channels. The parameters of the Gaussian fits (in dark blue) are in Table \ref{table_results}. Vertical lines mark the position of the central peak or its components.}
  \label{fig_4848_with_spectra}
\end{minipage}

\begin{minipage}{0.95\textwidth}
  \centering
   \includegraphics[width=\textwidth]{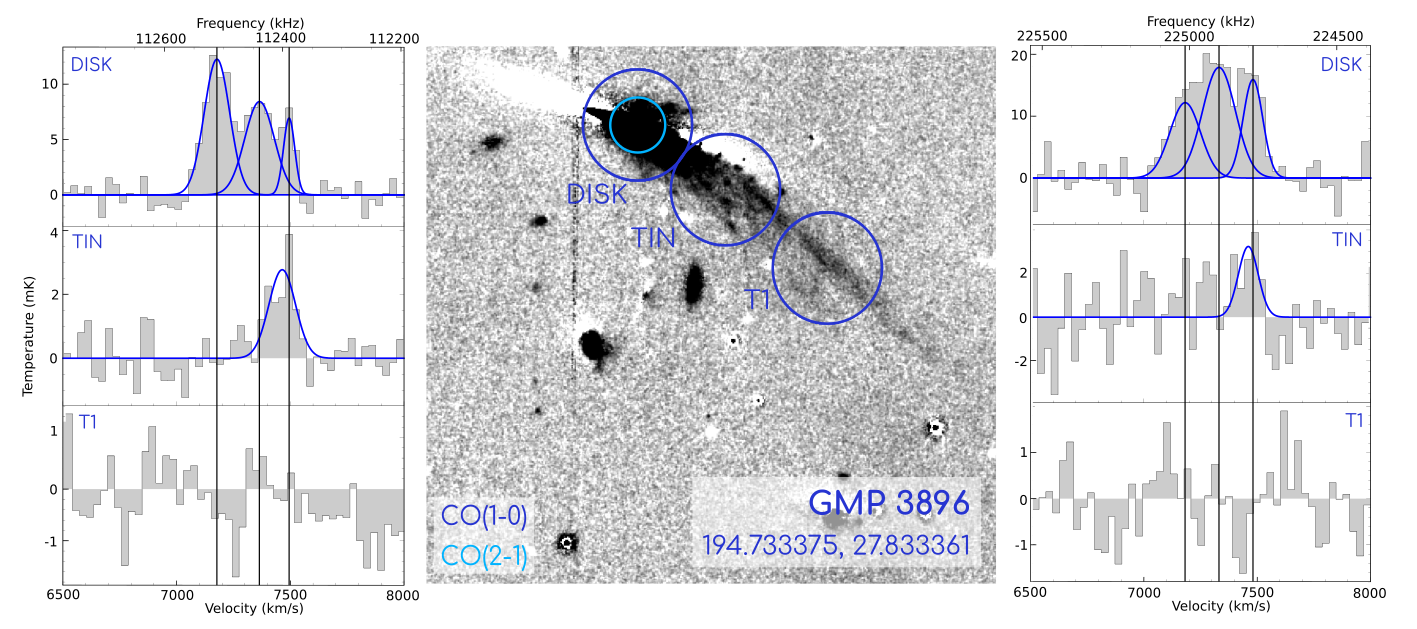}
   \captionof{figure}{Same as Fig. \ref{fig_4848_with_spectra} but for galaxy GMP~3896. }
   \label{fig_3896_with_spectra}
\end{minipage}

\begin{figure*}
   \centering
   \resizebox{0.9\hsize}{!}
   {\includegraphics[width=\hsize]{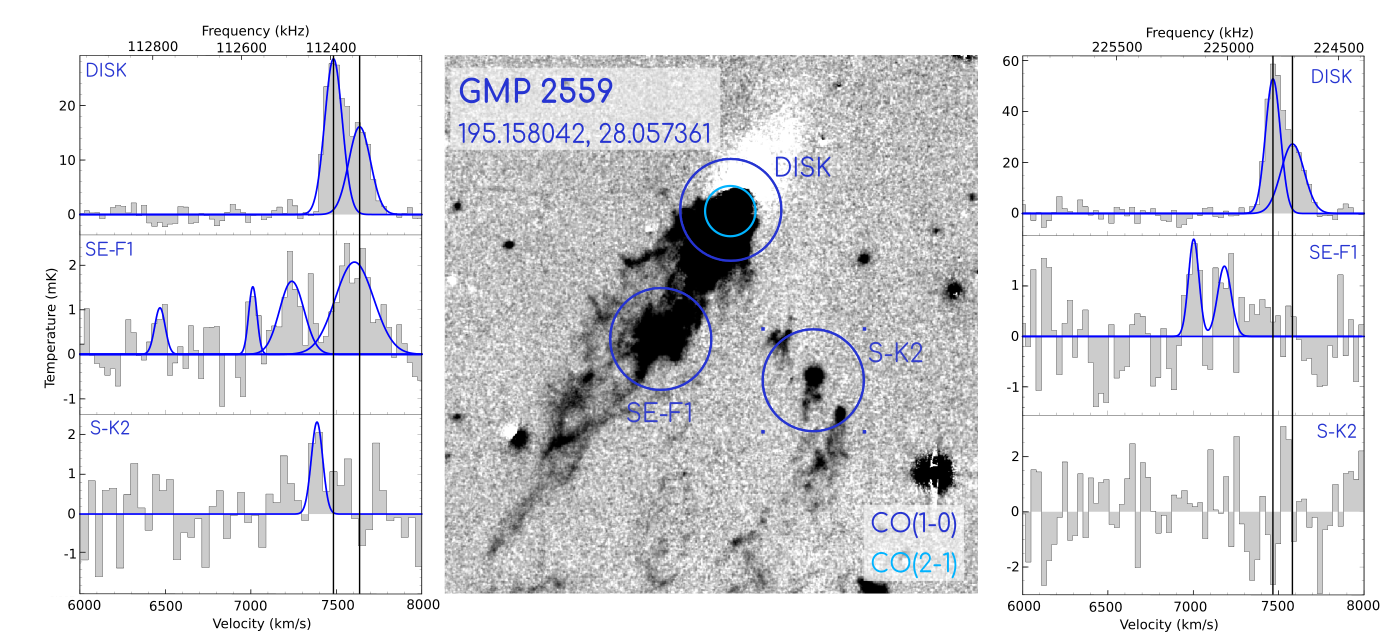}}
   \caption{Same as Fig. \ref{fig_4848_with_spectra} but for galaxy GMP~2559. }
   \label{fig_2559_with_spectra}
\end{figure*}
   
\begin{figure*}
   \centering
   \resizebox{0.9\hsize}{!}
   {\includegraphics[width=\textwidth]{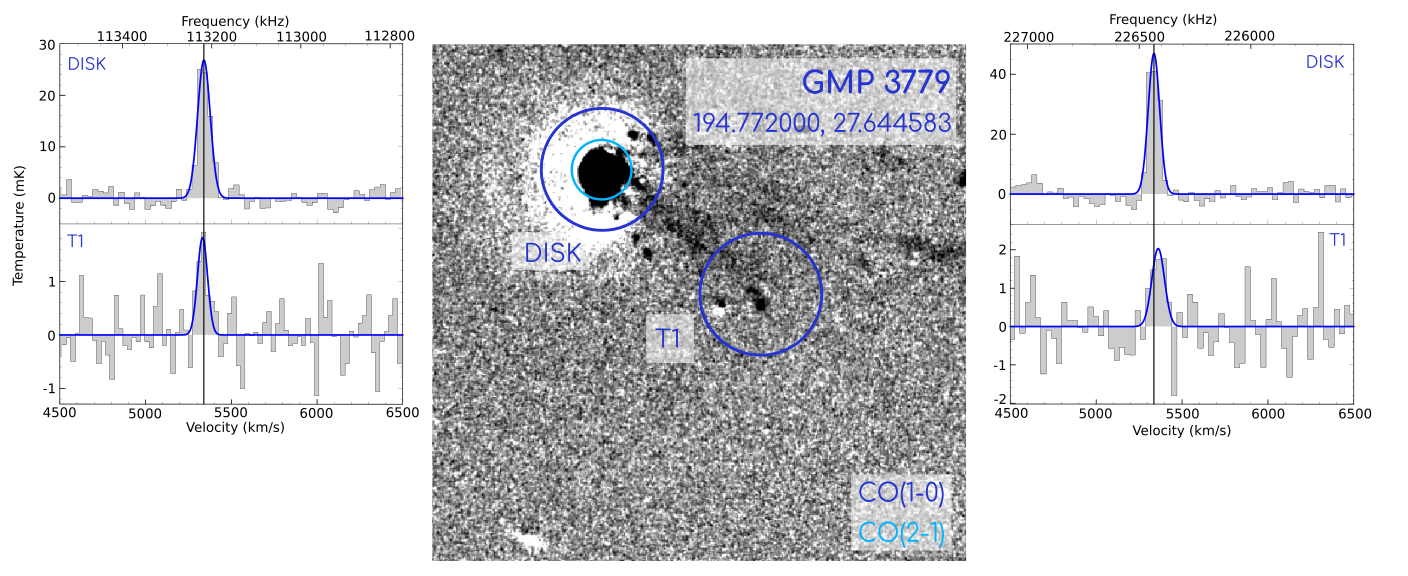}}
   \caption{Same as Fig. \ref{fig_4848_with_spectra} but for galaxy GMP~3779. }
   \label{fig_3779_with_spectra}
\end{figure*}
   
\begin{figure*}
    \centering
    \resizebox{0.9\hsize}{!}
    {\includegraphics[width=\textwidth]{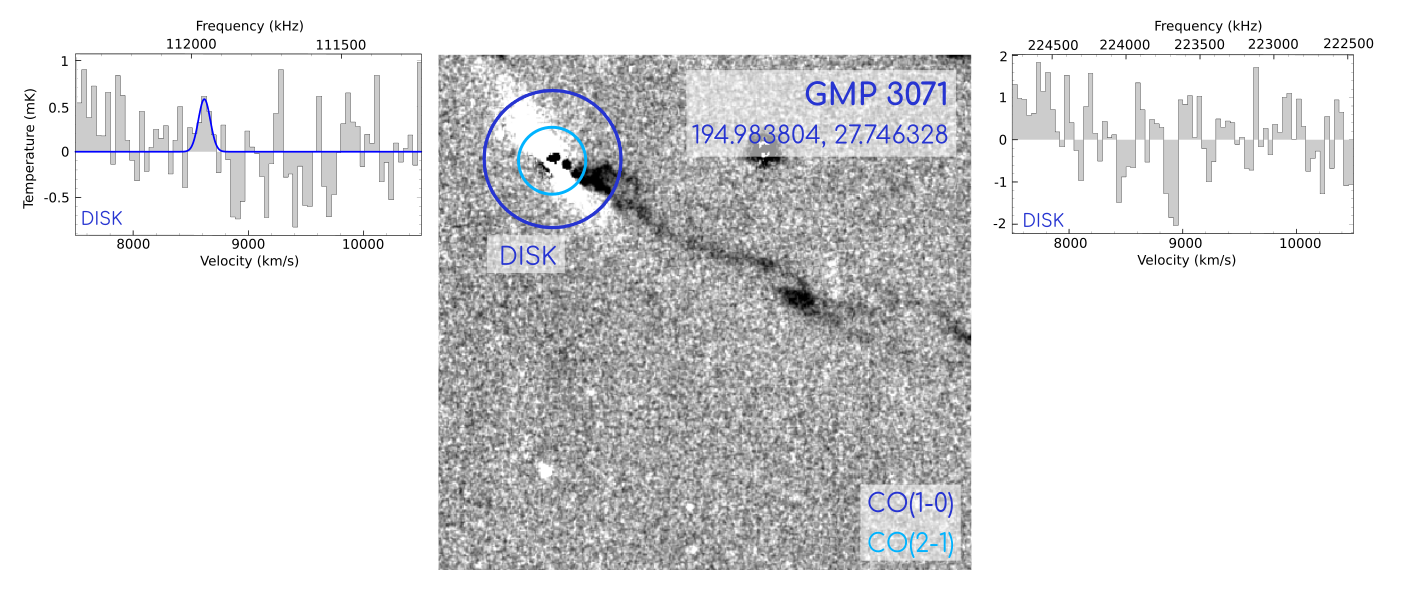}}
    \caption{Same as Fig. \ref{fig_4848_with_spectra} but for galaxy GMP~3071. }
    \label{fig_3071_with_spectra}
\end{figure*}
\clearpage

\section{Projected phase-space diagram}
\label{section_pps}
\begin{figure}
   \centering
   \includegraphics[width=\hsize]{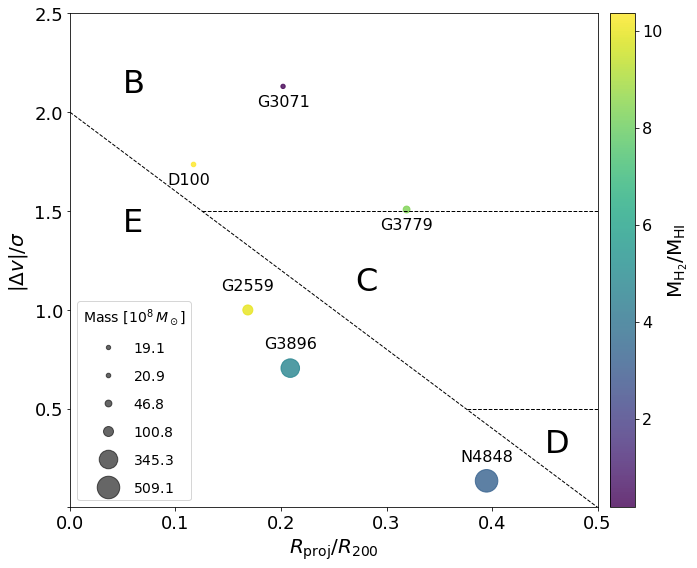}
      \caption{Projected phase-space diagram of our sample. Regions B-E reflect the time since infall of galaxies in them with region B being occupied by the most recent infallers, whereas region E hosts majority of ancient infallers (see Sect. \ref{section_history}). For context, we also show a stellar mass (marker size) and M$_{H_2}$/M$_{HI}$ (marker color).}
         \label{fig_pps}
   \end{figure}

Useful tool to study the evolution of galaxies within the cluster is a projected phase-space (PPS) diagram which aims to categorize individual objects by their time since infall inside $R_{vir}$ using their relative velocity and projected distance from the cluster center. Following \citet{Rhee2017}, we divided the PPS diagram to five regions (A-E) that are designed to maximize the fraction of galaxies belonging to a particular time since infall group in any region. The regions are defined as follows: region A is mostly occupied by first infallers (not fallen inside R$_{vir}$ yet), region B is dominated by recent infallers (0 < $\tau_{inf}$ < 3.6 Gyr), region D hosts mostly intermediate infallers (3.6 < $\tau_{inf}$ < 6.5 Gyr), and region E has highest fraction of ancient infallers (6.5 < $\tau_{inf}$ < 13.7 Gyr). Region C is a mixing area between regions B and D. For details, see \citet{Rhee2017}. As our sample only occupies regions R$_{proj}$/R$_{200}$ < 0.5 and $\lvert \Delta v \rvert$/$\sigma$ < 2.5, in Fig. \ref{fig_pps} we show a zoomed-in version of the original diagram that only includes regions B-E. For normalization, we used $R_{200}$ $\simeq$ 70 arcmin = 1.987 Mpc \citep{Churazov2021} and $\sigma$ = 873 km s$^{-1}$ \citep{Sohn2020}.\par
Galaxies GMP 3071, GMP 3779, and D100 belong to the region of recent infallers. Arguably, we see a good example of how ram pressure stripping acts on low-mass galaxies. Even though they, according to the PPS diagram, did not spend a long time in the cluster virial radius, they posses distinct stripping features - molecular-rich tail in the case of D100 and GMP 3779, while GMP 3071 seems almost fully stripped. Indeed, recent work \citep{Infurna2026} suggests that while in the case of massive galaxies RPS progresses gradually from outside in, lighter galaxies are stripped more abruptly. Rest of our sample - NGC 4848, GMP 3896, and GMP 2559 belong to ancient infallers group. Here we need to stress that the regions of a PPS diagram are made statistically and they do not necessarily reflect the real orbital history of a given object. Moreover, galaxies are placed onto the diagram based on projected distances and line-of-sight velocities while their true physical properties might be very different. Ancient infallers spent > 6.5 Gyr in the cluster virial radius and we do not believe that is the case for our galaxies. Typically, RPS acts on timescales of few Myr and as our sample showcases clear signs of active stripping, we assume they all entered the cluster recently in the terms of cited infall groups. The position of NGC 4848, GMP 2559, and GMP 3896 is presumably an effect of a selection process of our sample as we hand-picked galaxies with a prominent H$\alpha$ tail created by a strong RPS (and, hence, located close to the cluster center).

\end{appendix}

\end{document}